\documentstyle[11pt,paspconf,epsf]{article}
\begin{document}

\title{\Large \bf \center Feeding the central engine in giant radio galaxies}

\author{I. F. Mirabel$^{1,2}$ \& O. Laurent$^1$}
\affil{$^1$Service d'Astrophysique. Centre d'Etudes de Saclay. 91191 Gif/Yvette, 
France.\\
$^2$Instituto de Astronom\'\i a y F\'\i sica del Espacio. c.c. 67, suc 28. 
(1428) Buenos Aires, Argentina.}

\begin{abstract}

Giant radio galaxies are thought to be massive ellipticals powered by
accretion of interstellar matter onto a supermassive black
hole. Interactions with gas rich galaxies may provide the interstellar
matter to feed the active galactic nucleus (AGN). To power radio
lobes that extend up to distances of hundreds of kiloparsecs, gas has
to be funneled from kiloparsec size scales down to the AGN at rates 
of $\sim$1 M$_{\odot}$ yr$^{-1}$ during $\geq$10$^8$
years. Therefore, large and massive quasi-stable structures of gas and 
dust should exist in the deep interior of the giant elliptical hosts 
of double lobe
radio galaxies. Recent mid-infrared observations with ISO revealed 
for the first time a bisymmetric spiral structure with the
dimensions of a small galaxy at
the centre of Centaurus A (Mirabel et al. 1999). The spiral was
formed out of the tidal debris of accreted gas-rich object(s) and has
a dust morphology that is remarkably similar to that found in barred
spiral galaxies. The observations of the closest AGN to Earth suggest
that the dusty hosts of giant radio galaxies like CenA, are
``symbiotic" galaxies composed of a barred spiral inside an
elliptical, where the bar serves to funnel gas toward the AGN.
\end{abstract}



\section{A barred spiral at the centre of Centaurus A}

The mid-infrared and submillimeter observations of the dust emission
in CenA revealed a bisymmetric structure of 5$^\prime$ ($\sim$\,5
kpc for a distance of 3.5 Mpc) in total length (Mirabel et
al. 1999). Figure 1 shows that in contrast to the optical dark lanes
which show a wide and somewhat chaotic distribution, the structure of
the mid-infrared emission is remarkably thin, smooth and
bisymmetric. The emitting dust is less extended and displaced from the
most prominent optical dark lanes. This displacement is not due to
major differences between the spatial distributions of the cold and
very warm dust components. In a three dimensional tilted and warped
disk, projection effects play an important role, and the optical
appearance of the dark lanes in a luminous ellipsoidal system may be
affected by relatively small amounts of cold dust in the outer parts
of the bending disk that are located in the foreground side of the
luminous ellipsoidal distribution of stars.

\begin{figure}
\vspace{-1cm}
\plotone{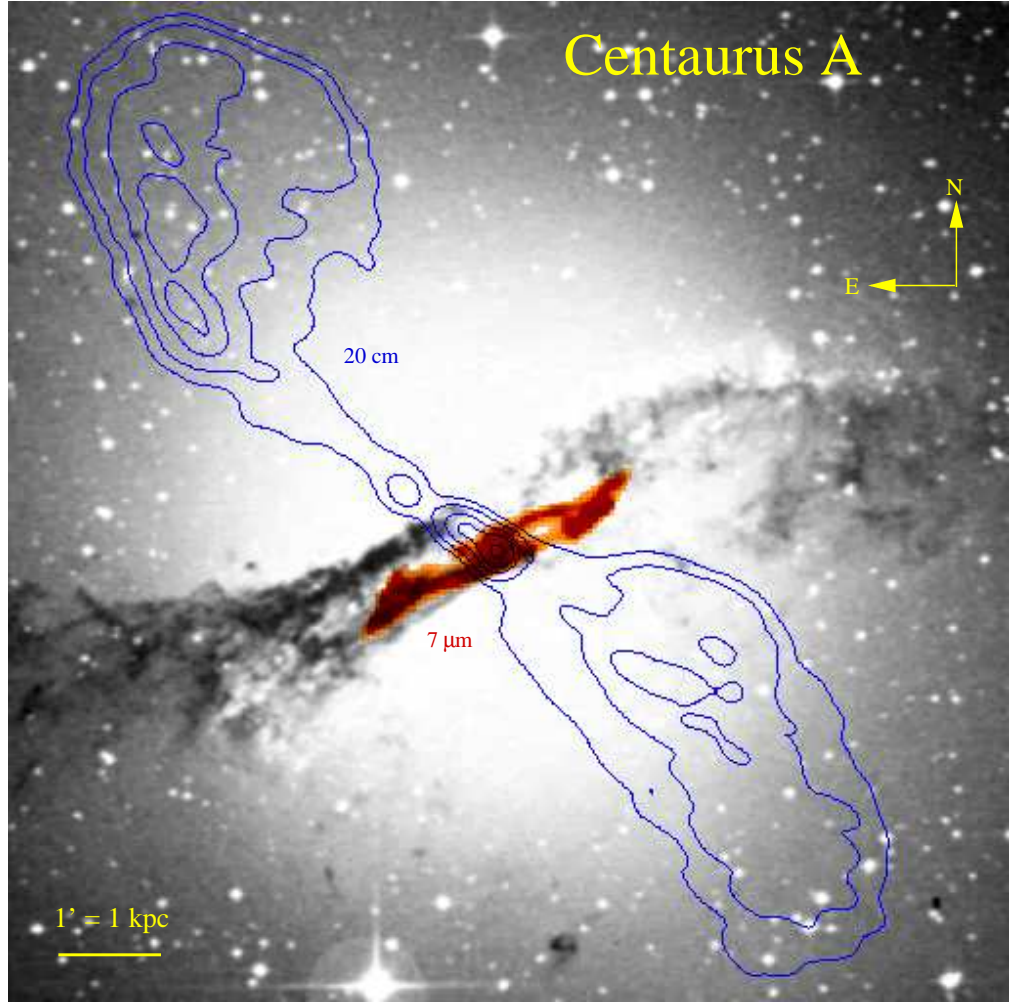}
\hspace{-4mm}
\caption
{\small{The ISO 7\,$\mu$m emission (dark structure) and
VLA 20 cm continuum in contours (Condon et al. 1996), overlaid on an
optical image from the Palomar Digital Sky Survey. The emission from
dust with a bisymmetric morphology at the centre is about 10 times
smaller than the overall size of the shell structure in the elliptical
(Malin et al. 1983) and lies on a plane that is almost parallel to the
minor axis of its giant host. Whereas the gas associated to the spiral
rotates with a maximum radial velocity of 250 km s$^{-1}$, the
ellipsoidal stellar component rotates slowly approximately
perpendicular to the dust lane (Wilkinson et al. 1986). The
synchrotron radio jets shown in this figure correspond to the inner
structure of a double lobe radio source that extends up to 5$^{\circ}$
($\sim$ 300 kpc) on the sky. The jets are believed to be powered by a
massive black hole located at the common dynamic center of the
elliptical and spiral structures.}}
\end{figure}

The structures seen in CenA with higher angular resolution by means of
the mid-infrared observations with ISO are inside the general
distribution of the molecular gas (Eckart et al. 1990; Rydbeck et
al. 1993). Figure 2 shows that the interpretation of the bisymmetric
structure at the centre of CenA as a barred spiral is fully consistent
with the morphology of the dust lanes observed in galaxies classified
as barred spirals, and the kinematics observed in CO
data. Furthermore, the interpretation of a barred structure is
consistent with theoretical models (Athanassoula, 1992) that predict
shocks at the leading edge of bars, producing an arc-like appearance
of the warm dust. The barred spiral is a dynamic instability, i.e. a
density wave in the warped disk of gas and dust.

\begin{figure}
\hspace{-3mm}
\plotone{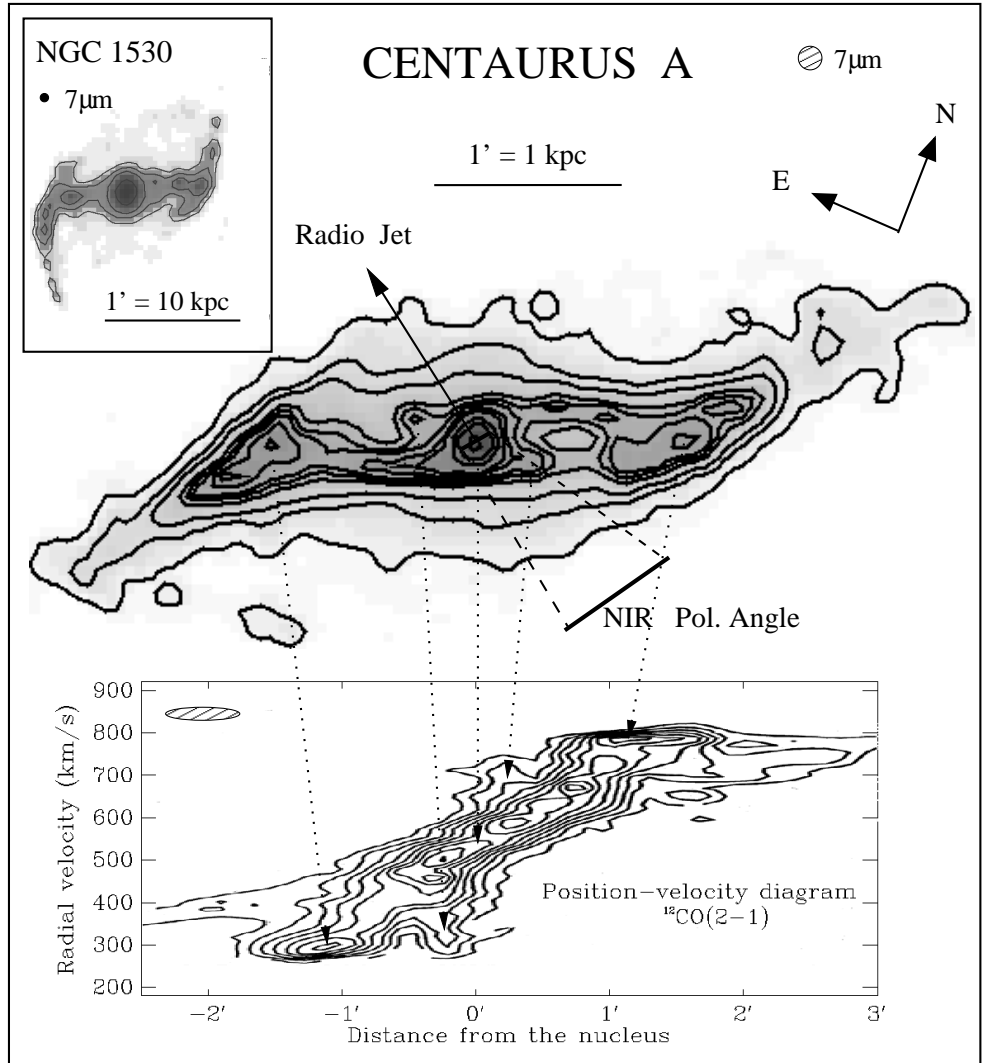}
\vspace{-4mm}
\caption
{\small{ISO 7\,$\mu$m image and $^{12}$CO(2-1) position-velocity map
(Quillen et al. 1992) of
the central region of CenA.  Note the similarity in morphology with
the 7\,$\mu$m image of the prototype barred spiral NGC\,1530. In
NGC\,1530 the plane that contains the bar is tilted by
$\sim$\,55$^{\circ}$ to the line of sight,
whereas in CenA it is tilted by $\sim$\,18$^{\circ}$.  The overall
structure exhibited by the 7\,$\mu$m emission from CenA is that of a
barred spiral with a primary bar extending $\sim$\,1$^\prime$ in
radius from the nucleus, connected in its outer ends to trailing
spiral arms. The strongest 7\,$\mu$m emission from the primary bar is
along its leading edge in what takes the form of two slightly curved
arcs, where shocks and density enhancements take place. The inner ends
of these two arcs are connected to what may be a secondary nuclear bar
whose position angle is defined by the NIR K(2.2\,$\mu$m) band
polarization (Packham et al. 1996).  On the plane of the sky the radio
jets appear perpendicular to the innermost polarization angle.  The
kinematics of the gas in the lower panel is consistent with a barred
spiral, where the bar rotates as a rigid body within
70$^{\prime\prime}$, whereas at radii larger than
$\pm$\,70$^{\prime\prime}$ the gas exhibits the differential rotation
(flat rotation curve) typical of galactic disks.}}
\end{figure}

ISOCAM could not resolve features smaller than $\sim$ 5\arcsec\ but
K(2.2\,$\mu$m) band polarization (Packham et al. 1996) due to the
passage of starlight through clouds that contain electrons and/or
aligned small grains can be used to trace with higher angular
resolution the preferential distribution of dust in the innermost
central region.  The position angle of the polarization in Figure 2
suggests the presence of a secondary bar, or ``nuclear'' bar of gas of
few hundred parsecs in size. This presumed secondary bar inside the
primary bar could be the dynamical instability that brings gas towards
the supermassive black hole, as proposed in the theoretical model by
Shlosman et al. (1989).  In fact, molecular gas absorption has been
detected in front of the compact nuclear source at millimeter
wavelengths, and it has been proposed that the absorption at
redshifted velocities represents gas falling into the centre (Israel
et al. 1991).

It is believed that the same accretion event(s) that began tearing
apart a gas-rich object(s) also created the faint stellar and gaseous
shells observed around CenA.  It is known that the accretion of a
small disk galaxy by a massive elliptical would lead to the complete
tidal disruption of the former as it spirals inward the potential
of the elliptical galaxy. In this process the gas decouples from the
stars and sinks more readily to the centre, forming a new disk out of
the gaseous component alone. In CenA the overall angular momentum of
the newly formed disk is not aligned with the major axis of the
elliptical. Therefore, the gaseous disk is subject to torques forcing
it to warp. Although gas is still settling towards the central
regions, the morphological and dynamical symmetry of the spiral
indicates that it is a stable structure and not a transient
feature. Rotating at 250 km, it must have undergone several full
rotations depending on how long after the initial encounter it took
the gas to settle into the central disk we now see.
 
Using near infrared photometry as well as the kinematics of the gas, a
disk-to-total mass ratio within the turnover radius of the rotation
curve of the order of $10^{-2}$ is obtained. N-body simulations would
rule out that a low-mass stellar bar has formed spontaneously in the
disk, and survived at steady-state.  However these simulations
consider stellar disks, while here we are dealing with a gaseous
one. Using the much lower velocity dispersion (typically 5-10 km
s$^{-1}$ in the gas, compared with 50 km s$^{-1}$ for a stellar disk
and 145 km s$^{-1}$ for the spheroidal component in CenA (Wilkinson et
al. 1986, Eckart et al. 1990) a much lower Toomre's Q parameter (of
the order of 1) is derived for the gaseous disk in CenA. Therefore, a
gaseous disk is much more self-gravitating than a stellar one of
similar mass, and the bar in CenA can be in a quasi-steady state and
might have formed spontaneously, or be driven.

\section{ The symbiotic galaxy Centaurus A: a template for giant radio galaxies}

Heckman et al. (1986) have shown that radio galaxies with giant radio lobes 
have peculiar optical morphologies. In this respect, 
CenA is not an exception and could serve as a well-positioned
template to examine in detail the clues to the origin and evolution of
activity in early-type radio galaxies with the same radio morphology,
namely, with giant double radio lobes. Prominent dust bands are
frequently observed in the hosts of this type of radio
galaxies. Fornax A, the second nearest giant radio galaxy after CenA
exhibits clear signs for the merger of gas-rich galaxies with a dusty
early type galaxy.  Cygnus A, the prototype radio galaxy with double
morphology is crossed by prominent optically dark bands that contain
$\sim$10$^8$ M$_{\odot}$ of dust (Robson et al. 1998). It has been
shown that in dusty radio galaxies with double radio structure, the
dust is usually found perpendicular to the radio axis (Kotani \&
Ekers 1979; van Dokkum \& Franx 1995) which suggests a connection
between the mechanism leading to these double radio morphologies and
the rotation axis of the dusty disk.

The specific mechanism in rapidly rotating disks of gas and dust that
brings fuel to the central engine in radio loud AGNs has been
difficult to probe observationally for several reasons. First,
galaxies like CenA are at greater distances (for instance, Fornax A is
5-10 times and Cygnus A is $\sim$ 70 times more distant than CenA),
and at those distances it is difficult to see the detailed morphology
of dust and gas on scales $\leq$ 100 pc. Second, at optical and
near-infrared wavelengths the light from the old stellar population
with a giant ellipsoidal distribution overwhelms any emission from
dust and newly formed stars in the deep interior. It is in the
mid-infrared that the emission from very warm dust can be better
traced, and to this end we had to wait for the unprecedented
capabilities of ISOCAM. Third, the observation of the cold gas
distribution by means of millimeter observations of weak molecular
line emission on top of the strong continuum of powerful radio
galaxies is a difficult task.

The need for the presence of bars in AGNs has been questioned because
the ocurrence of bars in Seyfert galaxies is not higher than in normal
galaxies (McLeod \& Riecke, 1995; Ho et al. 1997; Mulchaey \& Regan,
1997). Furthermore, using HST images in the optical and near-infrared,
no signatures of dust {\it absorption} with barred structures have
been found in Seyfert 2's (Regan \& Mulchaey, 1999). However, it is 
difficult to trace the presence of dust inside bulge and ellipsoidal 
systems using only optical and near-infrared observations because at 
those wavelengths the stellar emission dominates. 
Mirabel et al. (1999) have shown that the true structure of the dust in 
the deep interior of CenA can only be revealed by the emission at 
wavelengths longer than 3\,$\mu$m. In CenA the dust absorption derived 
from colors between the near-infrared and the optical (Quillen, 1993) 
only trace the foreground side of the structure shown in Figures 1 and 2.

It is believed that the AGN
activity in Seyfert galaxies is much more sporadic than in giant radio
galaxies and that Seyferts may require accretion masses with sizes 
that are orders of magnitude smaller than those needed 
to power the giant lobes of radio galaxies. The observation of 
{\it emission} from the accreeting fuel in the central regions of 
Seyfert galaxies may have to await the high angular resolution and 
sensitivity of the NGST and MMA/LSA.

The barred spiral at the centre of CenA has dimensions comparable to
that of the small Local Group galaxy Messier 33. It lies on a plane
that is almost parallel to the minor axis of the giant
elliptical. Whereas the spiral rotates with maximum radial velocities
of $\sim$\,250 km s$^{-1}$, the ellipsoidal stellar component seems
to rotate slowly (maximum line-of-sight velocity is $\sim$\,40 km
s$^{-1}$) approximately perpendicular to the dust lane. The genesis,
morphology, and dynamics of the spiral formed at the centre of CenA
are determined by the gravitational potential of the elliptical, much
as a usual spiral with its dark matter halo. On the other hand, the
AGN that powers the radio jets is fed by gas funneled to the center
via the bar structure of the spiral. The spatial co-existence and
intimate association between these two distinct and dissimilar systems
suggest a ``symbiotic" association.

\section{Conclusions} 

1) The observation of dust emission from the centre of CenA opens the more
general question on whether the hosts of giant radio galaxies are symbiotic
galaxies composed of spirals at the centre of giant ellipticals.
\\
2) The true structure of dust in the deep interior of ellipsoidal stellar 
systems is better traced by {\it emission} from dust at wavelengths longer 
than 3 microns, rather than by absorption using colors from near-infrared 
and optical photometry. 
\\
3) The accreting matter in Seyferts is likely to have masses and sizes 
that are several orders of magnitude smaller than in giant radio galaxies. 
To trace the accreting gas and dust in Seyferts, the angular resolution and 
sensitivity of the NGST and MMA/LSA are needed.

\end{document}